\documentstyle[epsf,axodraw]{aipproc}
\begin{document}
\begin{flushright}
$
\begin{array}{l}
\mbox{UB--HET--00--02}\\
\mbox{July~2000}\\[5.mm]
\end{array}
$
\end{flushright}
\title{Theoretical Challenges for a Precision Measurement of the $W$
Mass at Hadron Colliders\footnote{Talk given at the MRST Conference,
Rochester, NY, May~8~--~9, 2000, to appear in the Proceedings}} 
\author{U.~Baur}
\address{Department of Physics, State University of New York, Buffalo,
New York 14260} 
\maketitle
\begin{abstract}
{ We discuss the ${\cal
O}(\alpha)$ electroweak radiative corrections to $W$ and $Z$ boson 
production and their impact on the measurement of the $W$ mass at 
hadron colliders. The results of recent improved 
calculations are presented. We also briefly discuss the ${\cal O}(\alpha)$
corrections to Drell-Yan production in the high invariant mass region. }
\end{abstract}
 
\section{Introduction}
\label{sec:intro}

The Standard Model of electroweak interactions (SM)
so far has met all experimental challenges and is now tested at the 
$0.1\%$ level~\cite{lepewk}. However, there is little direct
experimental information on the mechanism  which generates the masses of
the weak gauge bosons. In the SM, spontaneous symmetry breaking is
responsible for mass generation. The existence of a Higgs boson
is a direct consequence of this mechanism. At present the negative
result of direct searches performed at LEP2 imposes a lower bound
of $M_H>107.9$~GeV~\cite{lephiggs} on the Higgs boson mass. Indirect
information on the mass of the Higgs boson can be extracted from the 
$M_H$ dependence of radiative corrections to the $W$ boson mass, $M_W$,
and the effective weak mixing angle, $\sin^2\theta^{lept}_{eff}$.
Assuming the SM to be valid, a global fit to
all available electroweak precision data yields a 95\%
confidence level upper limit on $M_H$ of 188~GeV~\cite{straes}. 

In order to extract more accurate information on $M_H$ from electroweak
data, it is very important to measure $M_W$ more precisely. Currently, 
the $W$ boson mass is 
known to $\pm 38$~MeV~\cite{straes} from direct measurements. Further 
improvement in the $W$ mass uncertainty is expected from this years
LEP~II data taking, and Run~II of the Tevatron~\cite{Tev2000} which is
scheduled to begin in March~2001. The ultimate precision
expected for $M_W$ from the combined LEP2 experiments is approximately 
35~MeV~\cite{LEPWmass}. At the Tevatron, integrated luminosities of
order 2~fb$^{-1}$ are foreseen for Run~II, and one
expects to measure the $W$ mass with a precision of approximately
40~MeV~\cite{Tev2000} per experiment and decay channel. Preliminary
studies indicate that measuring $M_W$ at the LHC with an with a
precision of 25~MeV~\cite{ewlhc} per experiment and decay channel should 
be possible, although very challenging.

In order to measure the $W$ boson mass with high
precision in a hadron collider environment, it is necessary to fully 
understand and control higher order QCD and electroweak (EW) corrections 
to $W$ production. The determination of the $W$ mass in a hadron 
collider environment
requires a simultaneous precision measurement of the $Z$ boson mass,
$M_Z$, and width, $\Gamma_Z$. When compared to the value measured at LEP, 
the two quantities help to accurately determine the energy scale and
resolution of the electromagnetic calorimeter, and to constrain the
muon momentum resolution~\cite{kotwal}. It is therefore also necessary
to understand the higher order EW corrections to $Z$ boson production in 
hadronic collisions. 

Accurate predictions for $W$ and $Z$ production which include the
complete ${\cal O}(\alpha)$ corrections are also needed for a
measurement of the $W$ cross section, the $W/Z$ cross section ratio, 
the determination of the $W$
width, and for extracting $\sin^2\theta^{lept}_{eff}$ from the forward
backward asymmetry in the $Z$ peak region. In addition, precise
calculations of the Drell-Yan cross section are needed in searches for new 
physics beyond the SM, such as hidden extra dimensions or additional $Z$ 
bosons. 

In a previous calculation of the EW radiative corrections to $W$ and $Z$ 
production, only the final state photonic corrections were correctly 
included~\cite{BK}. The sum of the soft and virtual parts was estimated
from the inclusive ${\cal O}(\alpha^2)$ $W\to\ell\nu(\gamma)$ and
$Z\to\ell^+\ell^-(\gamma)$ ($\ell=e,\,\mu$) 
width and the hard photon bremsstrahlung contribution. Initial state, 
interference, and weak contributions to the ${\cal O}(\alpha)$ 
corrections were ignored altogether. The unknown part of the ${\cal 
O}(\alpha)$ EW corrections in Ref.~\cite{BK}, combined with effects 
of multiple photon emission, have been estimated to contribute a 
systematic uncertainty of $\delta M_W=15-20$~MeV to the measurement of 
the $W$ mass in Run~I~\cite{kotwal}. Clearly, in order to achieve the 
accuracies envisioned for Run~II and the LHC, improved theoretical 
calculations are required. 

Recently, new and more accurate calculations of the ${\cal
O}(\alpha)$ EW corrections to $W$~\cite{BKW} and $Z$ boson production in 
hadronic collisions~\cite{BKS,BBHSW} became available. In this talk I
present an overview of these calculations. They include most of the
contributions which were previously ignored.
In Section~\ref{sec:two} I briefly describe the calculation of the
${\cal O}(\alpha)$ EW corrections to $Z$ boson and high mass Drell-Yan
production. In Section~\ref{sec:three} I summarize the results of
Ref.~\cite{BKW}. In Section~\ref{sec:four} I present a brief summary and 
outlook. 

\section{Electroweak Corrections to $Z$ boson and High Mass Drell-Yan
Production}
\label{sec:two}

\subsection{The $\boldmath{{\cal O}(\alpha)}$ QED Corrections to
Di-lepton Production}

For $p\,p\hskip-7pt\hbox{$^{^{(\!-\!)}}$} \to
Z,\gamma^*\to\ell^+\ell^-$, the pure QED corrections form a separately
gauge invariant set of diagrams. The first step towards a full calculation 
of the ${\cal O}(\alpha)$ corrections to $Z$ boson production thus
consists of performing a
calculation of the pure QED corrections. The diagrams contributing to
the ${\cal O}(\alpha)$ QED corrections can be separated into gauge 
invariant subsets corresponding to initial and final state corrections. 

To perform the calculation, a Monte Carlo method for
next-to-leading-order (NLO) calculations similar to that described in 
Ref.~\cite{NLOMC} was used. With the Monte Carlo method, it is easy to 
calculate a variety of observables simultaneously and to simulate detector 
response. The collinear singularities associated with final state photon 
radiation are regulated by the mass of the leptons. The associated
mass singular logarithms of the form $\log(\hat s/m_\ell^2)$, where $\hat
s$ is the squared parton center of mass energy and $m_\ell$ is the
charged lepton mass, are included in our calculation, but the very small
terms of ${\cal O}(m_\ell^2/\hat s)$ are neglected. 

The collinear singularities associated with initial state
photon radiation can be removed by universal collinear counter terms
generated by ``renormalizing'' the parton distribution
functions (PDF's)~\cite{spies}, in complete analogy to gluon emission in
QCD. In addition to the collinear counterterms, finite terms can be absorbed
into the PDF's, introducing a QED factorization scheme dependence. We
have carried out our calculation in the QED $DIS$ and QED $\overline{MS}$
scheme. In order to treat the ${\cal O}(\alpha)$ initial state QED 
corrections to $Z$ boson production in hadronic collisions in a
consistent way, QED corrections should be incorporated in the global 
fitting of the PDF's using the same factorization scheme which has been
employed to calculate the cross section. Current fits to the PDF's do
not include QED corrections, which introduces a small uncertainty into
the calculation. 

\begin{figure}
\begin{center}
\begin{tabular}{cc}
\epsfysize=2.in \epsffile{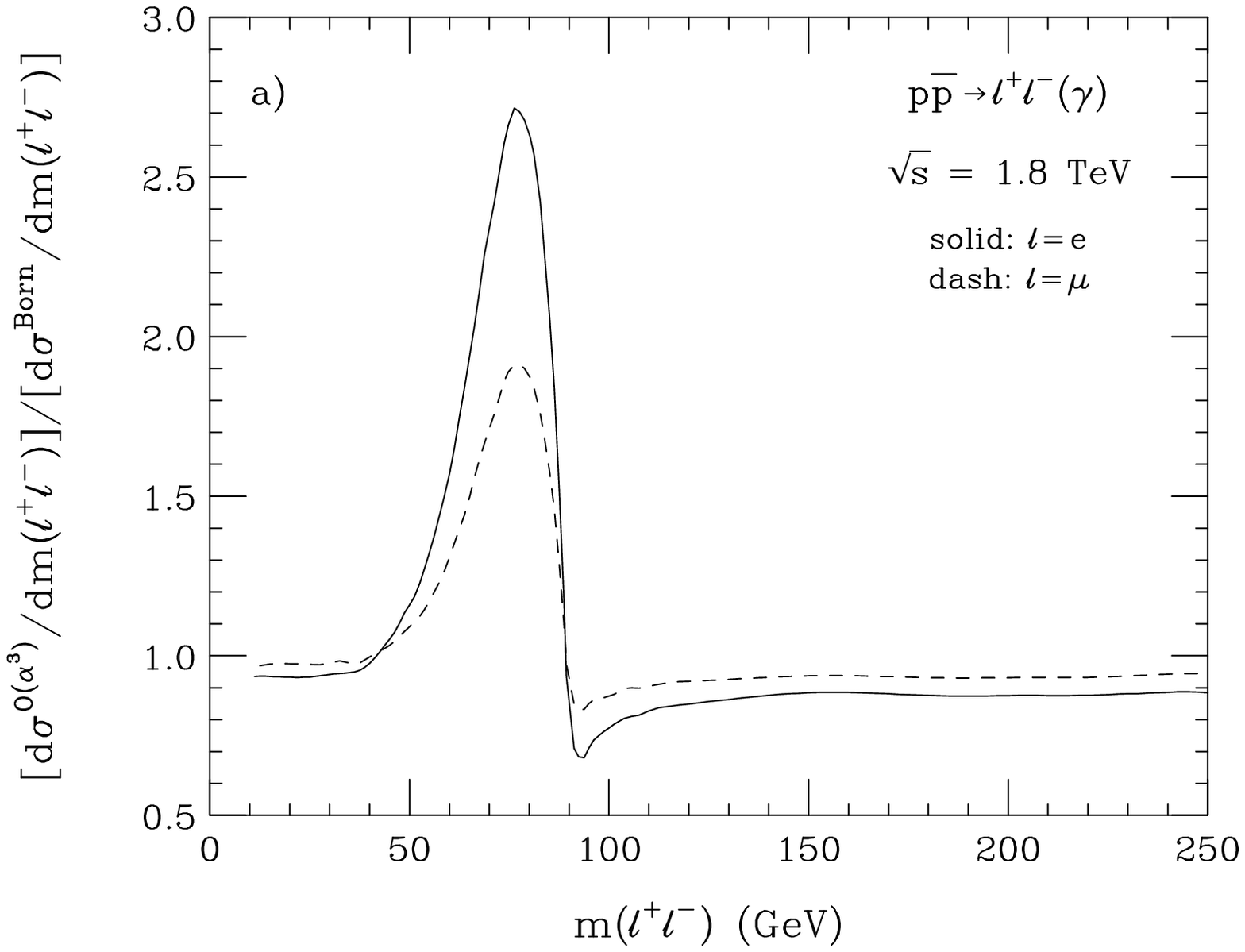} & 
\epsfysize=2.in \epsffile{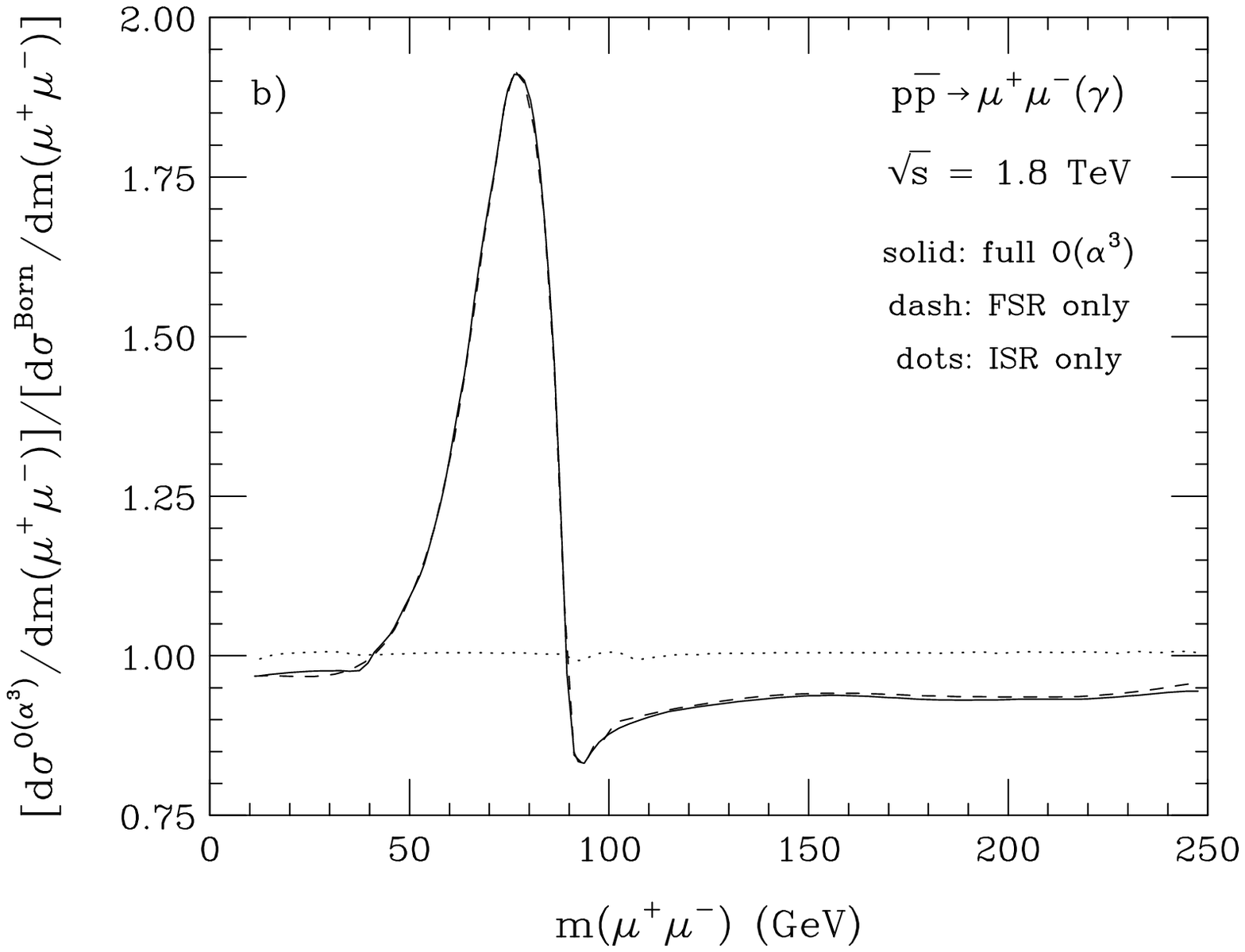} \\ 
\end{tabular}
\end{center}
\caption{Ratio of the \protect{${\cal O}(\alpha^3)$} and lowest order
differential cross sections as a function of the di-lepton invariant
mass for $p\bar p\to\ell^+\ell^-(\gamma)$ at
$\protect{\sqrt{s}=1.8}$~TeV. Part a) of the figure shows the ratios for 
electrons and muons in the final state. Part b) shows, for $p\bar
p\to\mu^+\mu^-(\gamma)$, the result for the
full set of \protect{${\cal O}(\alpha^3)$} QED diagrams (solid line), and
for taking only final state (dashed line) and initial state corrections
(dotted line) into account. }
\label{fig:one}
\end{figure}

In Fig.~\ref{fig:one}a we display
the ratio of the ${\cal O}(\alpha^3)$ and the Born cross section as a 
function of the $\ell^+\ell^-$ invariant mass in $p\bar p\to\gamma^*,\,
Z\to\ell^+\ell^-$ at Tevatron energies. In the region
$40~{\rm GeV}<m(\ell^+\ell^-)<110$~GeV, the cross section ratio is seen
to vary rapidly. 
Below the $Z$ peak, QED corrections enhance the cross section by up to a 
factor~2.7 (1.9) for electrons (muons). The maximum enhancement of the 
cross section occurs at $m(\ell^+\ell^-)\approx 75$~GeV. At the $Z$ peak, the
differential cross section is reduced by about $30\%$ ($20\%$). For 
$m(\ell^+\ell^-)>130$~GeV, the ${\cal O}(\alpha)$ QED corrections 
uniformly reduce 
the differential cross section by about $12\%$ in the electron case, and 
$\approx 7\%$ in the muon case. 

In Fig.~\ref{fig:one}b, we compare the impact of the full ${\cal 
O}(\alpha)$ QED corrections (solid line) on the muon pair invariant 
mass spectrum with that of final state (dashed line) and initial state
radiative corrections (dotted line) only. Qualitatively similar results 
are obtained
in the electron case. Final state radiative corrections are seen to
completely dominate over the entire mass range considered. They are
responsible for the strong modification of the 
di-lepton invariant mass distribution. In contrast, initial state
corrections are uniform and small ($\approx +0.4\%$). 

The results shown in Fig.~\ref{fig:one} can be understood by recalling
that final state photon radiation leads to
corrections which are proportional to $\alpha\log(\hat s/m_\ell^2)$. These
terms are large, and significantly influence the shape of the di-lepton 
invariant mass distribution. Photon radiation from one of the leptons 
lowers the di-lepton invariant
mass. Events from the $Z$ peak region therefore are shifted towards 
smaller values of $m(\ell^+\ell^-)$, thus reducing the
cross section in and above the peak region, and increasing the rate
below the $Z$ pole. Due to the $\log(\hat s/m_\ell^2)$ factor, the
effect of the corrections is larger in the electron case. 

In Fig.~\ref{fig:one}, we have not taken into account realistic lepton
identification requirements. To simulate detector acceptance, we
now impose the following lepton transverse momentum ($p_T$) and rapidity
($\eta$) cuts, which are similar to those used by the CDF Collaboration
in Run~I: 
\begin{eqnarray}
p_T(e)>20~{\rm GeV,} & \qquad & |\eta(e)|<2.4, \\
p_T(\mu)>25~{\rm GeV,} & \qquad & |\eta(e)|<1.0.
\end{eqnarray}
In addition at least one of the electrons (muons) is required to have
$|\eta(e)|<1.1$ ($|\eta(\mu)|<0.6$). Uncertainties in the energy 
measurements of the charged leptons in the detector are simulated in 
the calculation by Gaussian smearing of the particle four-momentum
vector according to the CDF electron and muon momentum resolutions. 
The granularity of the detector and the size of the electromagnetic
showers in the calorimeter make it difficult to discriminate between 
electrons and photons with a small opening angle. One therefore
recombines the four-momentum vectors of the electron 
and photon to an effective electron four-momentum vector if both
traverse the same calorimeter cell. Muons are identified in a hadron 
collider detector by hits in the muon
chambers. In addition, one requires that the associated track is
consistent with a minimum ionizing particle. This limits the energy of a
photon which traverses the same calorimeter cell as the muon to be 
smaller than a critical value $E^\gamma_c$. In the subsequent
discussion, we assume $E^\gamma_c=2$~GeV. 

\begin{figure}
\begin{center}
\begin{tabular}{cc}
\epsfysize=2.in \epsffile{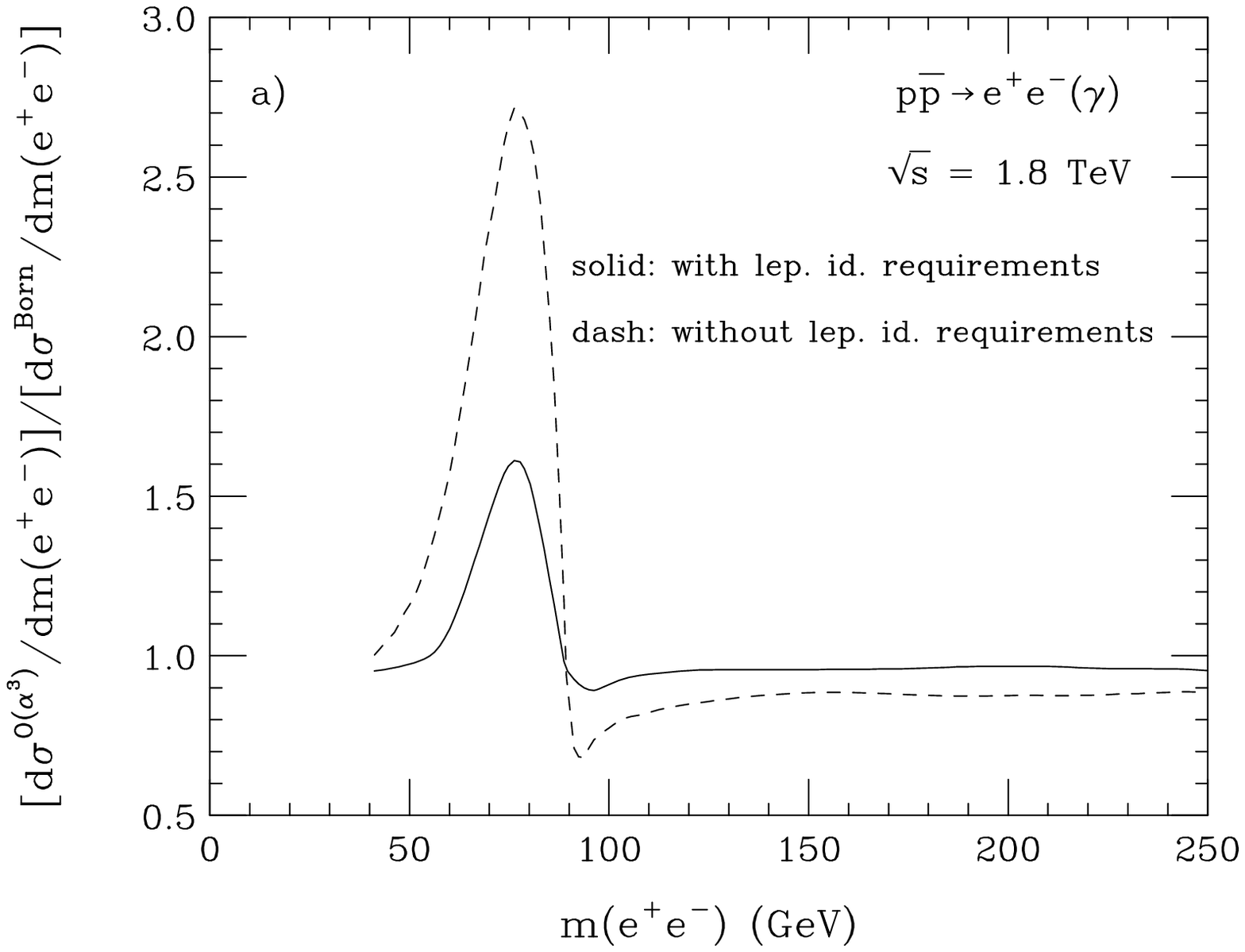} & 
\epsfysize=2.in \epsffile{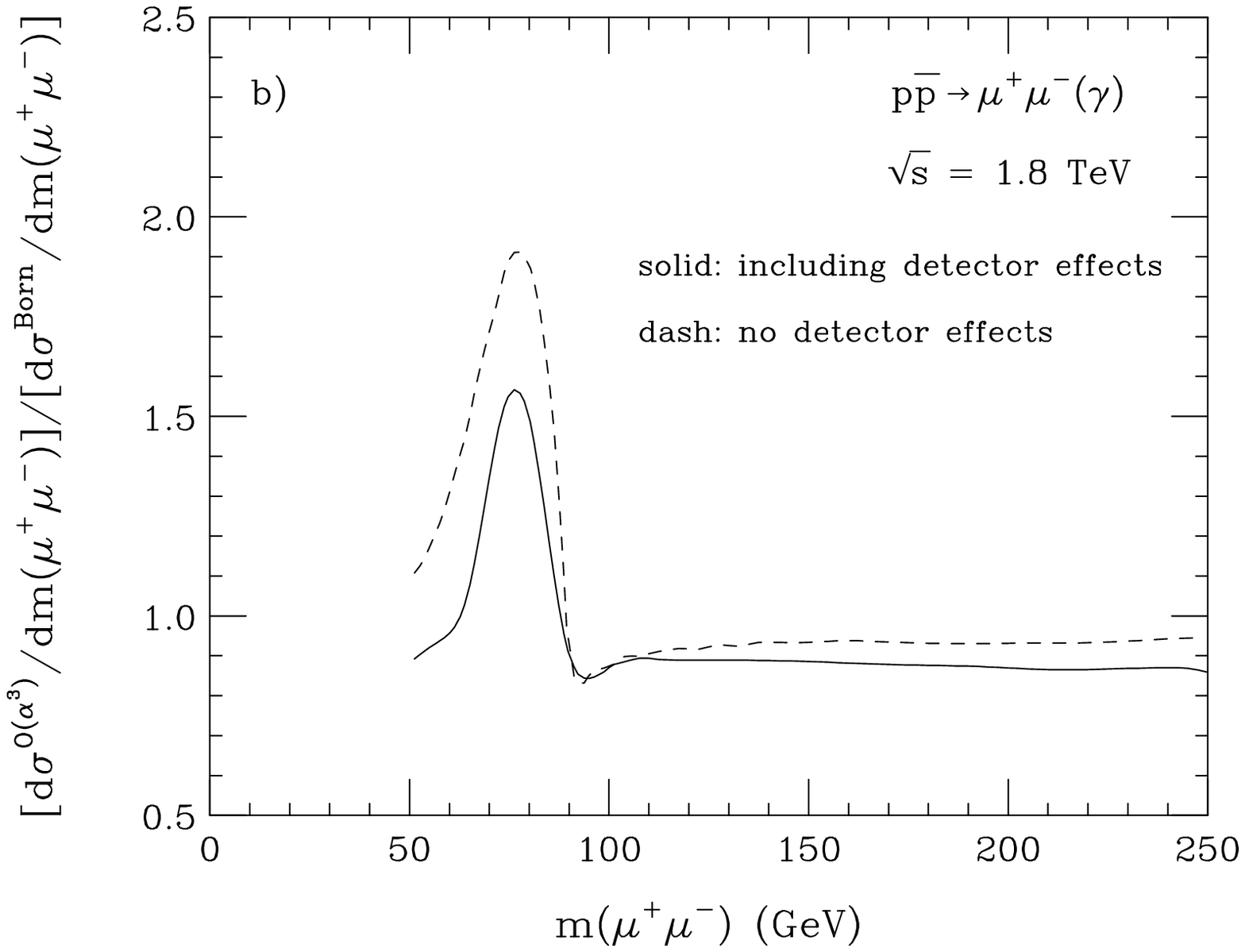} \\ 
\end{tabular}
\end{center}
\caption{Ratio of the \protect{${\cal O}(\alpha^3)$} and lowest order
differential cross sections as a function of the di-lepton invariant 
mass for a) 
$p\bar p\to e^+e^-(\gamma)$ and b) $p\bar p\to\mu^+\mu^-(\gamma)$ at 
$\protect{\sqrt{s}=1.8}$~TeV. The solid (dashed) lines show the cross
section ratio with (without) the detector effects described in the text.}
\label{fig:two}
\end{figure}
In Fig.~\ref{fig:two}a (Fig.~\ref{fig:two}b) we show how detector
effects change the
ratio of the ${\cal O}(\alpha^3)$ to leading order differential cross
sections as a function of the $e^+e^-$ ($\mu^+\mu^-$) invariant mass for 
$p\bar p$ collisions at $\sqrt{s}=1.8$~TeV. 
The finite energy resolution and the acceptance cuts have only a small
effect on the cross section ratio. The lepton identification criteria,
on the other hand, are found to have a large impact. Recombining the electron
and photon four-momentum vectors if they traverse the same calorimeter
cell eliminates the mass singular terms originating from final state
photon radiation. Although the recombination of the 
electron and photon momenta reduces the effect of the ${\cal O}(\alpha)$ 
QED corrections, the remaining corrections are still sizeable. Below
(at) the $Z$ peak, they enhance (suppress) the lowest order $e^+e^-$
differential cross section by up to a factor~1.6 (0.9) (see 
Fig.~\ref{fig:two}a). 
For muon final states (see Fig.~\ref{fig:two}b), the requirement of 
$E_\gamma<E^\gamma_c=2$~GeV for a photon which traverses the same 
calorimeter cell as the muon reduces the hard photon part of the 
${\cal O}(\alpha^3)$ $\mu^+\mu^-(\gamma)$ cross section. As a result,
the magnitude of the QED corrections below the $Z$ peak is reduced. 
At the $Z$ pole the corrections remain unchanged, and
for $\mu^+\mu^-$ masses larger than $M_Z$ they become more pronounced. 

From Figs.~\ref{fig:one} and~\ref{fig:two} it is clear that final state 
bremsstrahlung 
severely distorts the Breit-Wigner shape of the $Z$ resonance curve. As
a result, QED corrections must be included when the $Z$ boson mass is
extracted from data, otherwise the mass extracted is shifted to a lower
value. In the approximate treatment of
the QED corrections to $Z$ boson production used so far by the Tevatron 
experiments, only final state corrections are taken into account, and the 
effects of soft and virtual corrections are estimated from the inclusive 
${\cal O}(\alpha^2)$ $Z\to\ell^+\ell^-(\gamma)$ width and the hard photon 
bremsstrahlung contribution~\cite{BK}. When detector effects are taken
into account, the approximate calculation leads to a shift of the $Z$
mass of about $-150$~MeV in the electron case, and approximately 
$-300$~MeV in the muon case~\cite{kotwal}. The $Z$ boson mass extracted 
from the $\ell^+\ell^-$ invariant mass distribution which includes the
full ${\cal O}(\alpha^3)$ QED corrections
is found to be about 10~MeV smaller than that obtained using the 
approximate calculation of Ref.~\cite{BK}. 

The bulk of the shift in $M_Z$ originates from final state photon
radiation. This raises the question of how strongly multiple photon
radiation influences the measured $Z$ boson mass. An explicit
calculation of $\ell^+\ell^-\gamma\gamma$ production in hadronic
collisions~\cite{BS} shows that two photon radiation has a significant
impact on the shape of the $Z$ resonance curve. In order to determine
its effect on $M_Z$, more detailed simulations have to be carried out. 

\subsection{Including Weak Corrections}

So far, the purely weak corrections, which mainly consist of vertex
corrections and box diagrams with two massive bosons exchanges,
were ignored in our discussion. In the $Z$ peak region, these
corrections are small. However, at large energies, the effect of the 
weak vertex and box diagrams becomes large~\cite{log}. In this section
we present preliminary results of a new calculation~\cite{BBHSW} which
takes into account the purely weak corrections in $Z$ and Drell-Yan
production. 

\begin{figure}
\begin{center}
\begin{tabular}{cc}
\epsfysize=1.9in \epsffile{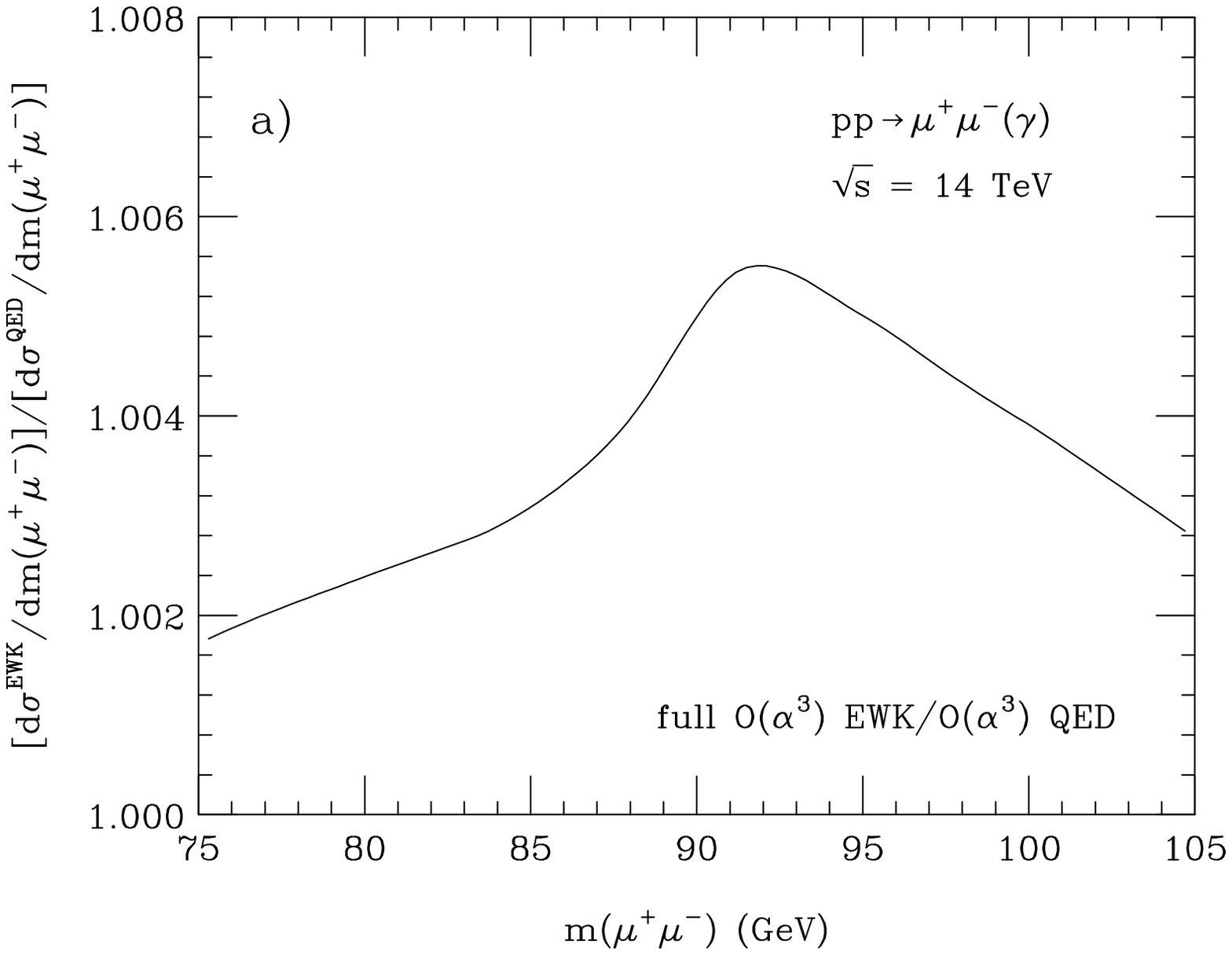} & 
\epsfysize=1.9in \epsffile{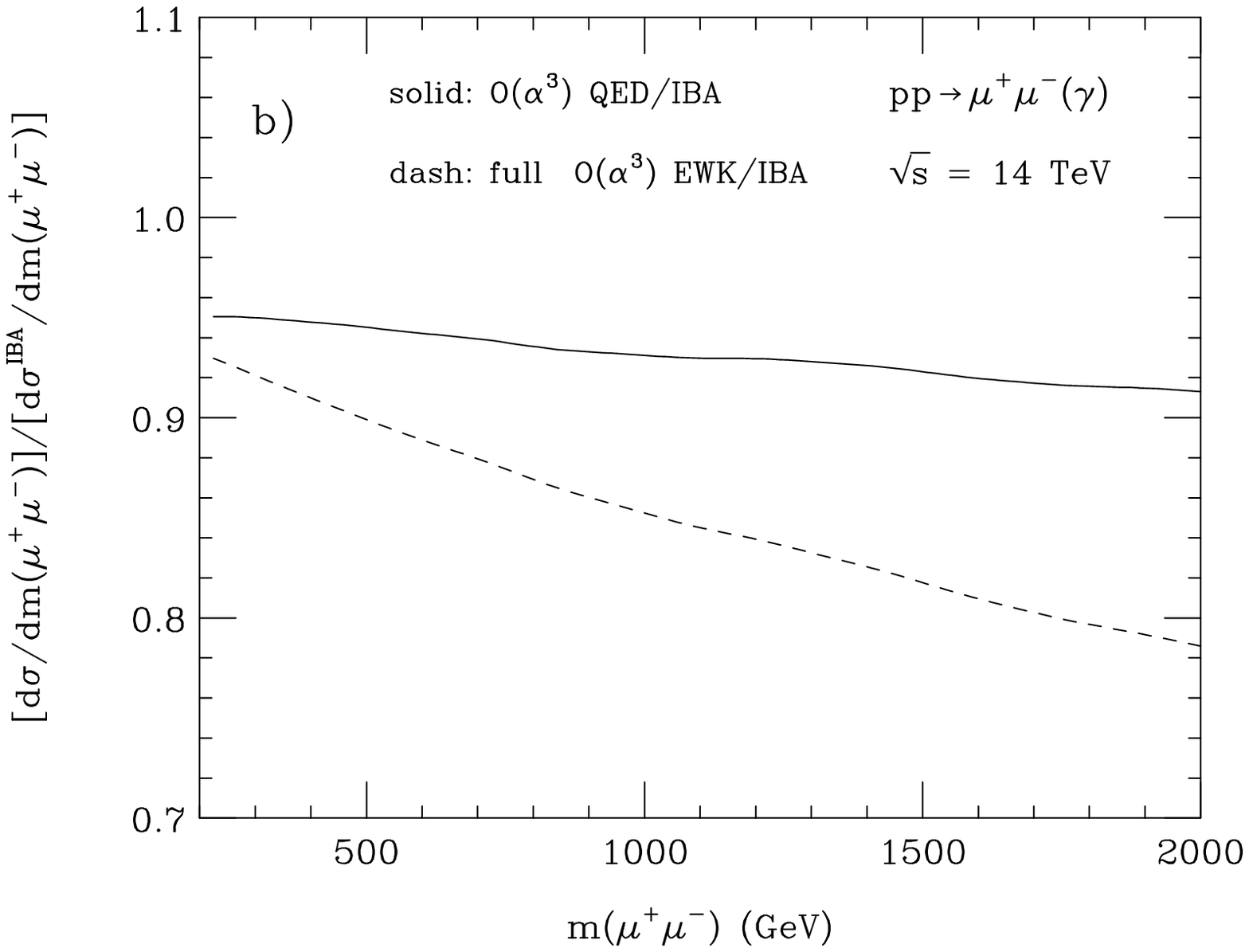} \\ 
\end{tabular}
\end{center}
\caption{a) Ratio of the full \protect{${\cal O}(\alpha^3)$}
electroweak  and
the \protect{${\cal O}(\alpha^3)$} QED differential cross sections in
the vicinity of the $Z$ pole at the LHC. b) Differential cross section
ratios at the LHC, displaying the size of the full \protect{${\cal 
O}(\alpha^3)$} electroweak and the \protect{${\cal O}(\alpha^3)$} QED 
corrections for large values of $m(\mu^+\mu^-)$. The cuts imposed are 
described in the text.}
\label{fig:three}
\end{figure}

In Fig.~\ref{fig:three}a we show the ratio of the full ${\cal O}(\alpha^3)$ 
electroweak and the ${\cal O}(\alpha^3)$ QED differential cross sections 
for $pp\to\mu^+\mu^-(\gamma)$ at the LHC as a function of the $\mu^+\mu^-$ 
invariant mass~\cite{ewlhc,BBHSW}. Here we have imposed a 
$p_T(\mu)>20$~GeV and a $|\eta(\mu)|<3.2$ cut, and used the improved
Born approximation (IBA) to evaluate the lowest order
contribution to the ${\cal O}(\alpha^3)$ QED cross section. Similar
results are obtained for the $e^+e^-$ final state and $p\bar p$ 
collisions at Tevatron energies.
The IBA incorporates the running electromagnetic coupling constant, the 
$Z$ propagator expressed in terms of the Fermi constant, $G_\mu$, and the
$Z$ boson mass and width measured at LEP, and the vector and axial
vector couplings expressed in terms of the effective leptonic weak
mixing angle. The ratio shown in Fig.~\ref{fig:three}a
directly displays the effect of the weak box diagrams and the
energy dependence of the weak coupling form factors. While the additional
weak contributions change the differential cross section by 0.6\% at
most, they do modify the shape of the $Z$ resonance curve.

Figure~\ref{fig:three}b compares the effect of the ${\cal O}(\alpha^3)$ QED 
corrections and the full ${\cal O}(\alpha^3)$ electroweak corrections
on the di-muon invariant mass distribution at the LHC for
$m(\mu^+\mu^-)$ values 
between 200~GeV and 2~TeV. Due to the presence of logarithms of the form
$\log(\hat s/M_Z^2)$, the weak
corrections become significantly larger than the QED corrections at
large values of $m(\mu^+\mu^-)$, and, eventually, may have to be
resummed~\cite{KPS}. For $m(\mu^+\mu^-)=2$~TeV, the full ${\cal
O}(\alpha^3)$ electroweak corrections are found to reduce the
differential cross section by more than 20\%.

\section{Electroweak Corrections to $W$ Boson Production}
\label{sec:three}

The calculation of the ${\cal O}(\alpha)$ corrections to $W$ boson
production~\cite{BKW} employs the same Monte Carlo method which was used 
in the $Z$ case. The collinear
singularities originating from initial state photon radiation are
again removed by counter terms generated by renormalizing the PDF's. 
Calculating the EW radiative corrections to $W$ boson
production, the problem arises how an unstable charged gauge boson can
be treated consistently in the framework of perturbation theory.
This problem has been studied in Ref.~\cite{dw} with 
particular emphasis on finding a gauge invariant decomposition of the 
EW ${\cal O}(\alpha)$ corrections into a QED-like and a modified weak part. 
In $W$ production, the Feynman diagrams which involve
a virtual photon do not represent a gauge invariant subset. 
In Ref.~\cite{dw} it was
demonstrated how gauge invariant contributions that contain the
infrared (IR) singular terms can be extracted from the virtual photonic 
corrections. These contributions can be combined with the also IR-singular 
real photon corrections in the soft photon region to form IR-finite 
gauge invariant QED-like contributions
corresponding to initial state, final state and interference
corrections. The IR finite remainder of the virtual photonic corrections 
and the pure weak one-loop corrections can be combined to separately
gauge invariant modified weak contributions to the $W$ boson
production and decay processes. 

\begin{figure}[t]
\centerline{
\epsfysize=2.9in
\epsffile{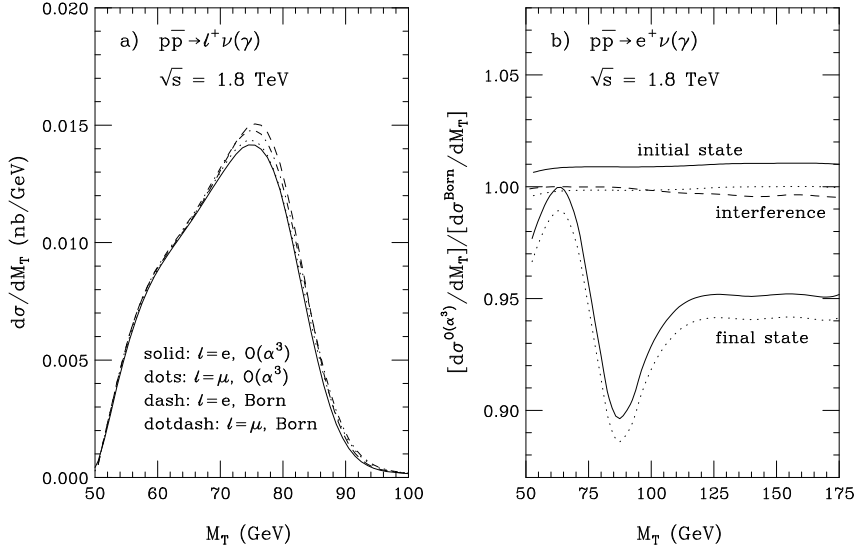}
}
\caption{a) The $M_T$ distribution for $p\bar p\to \ell^+\nu(\gamma)$ at
$\protect{\sqrt{s}=1.8}$~TeV at the Born level and including the EW
${\cal O}(\alpha)$ corrections. 
b) Ratio of the ${\cal O}(\alpha^3)$ and lowest order 
cross sections as a function of the transverse mass for $p\bar p\to
e^+\nu(\gamma)$ at
$\protect{\sqrt{s}=1.8}$~TeV for various individual contributions. The 
upper (lower) solid lines show the result for the QED-like initial 
(final) state corrections. The upper (lower) dotted lines give the cross
section ratios if
both the QED-like and modified weak initial (final) state corrections are 
included. The dashed lines display the result if only the 
initial -- final state interference contributions are included.}
\label{fig:four}
\end{figure}

Since hadron collider detectors cannot directly 
detect the neutrinos produced in the leptonic $W$ boson decays,
$W\to\ell\nu$, and cannot measure the
longitudinal component of the recoil momentum, there is insufficient
information to reconstruct the invariant mass of the $W$ boson. 
Instead, the transverse mass ($M_T$) distribution of the $\ell\nu$
system, or the $p_T$ distribution of the charged lepton, are 
used to extract $M_W$. The $M_T$ distribution for electron and muon
final states at the Born level and including ${\cal O}(\alpha)$
corrections at the Tevatron is shown in Fig.~\ref{fig:four}a. The
various individual contributions to the 
EW ${\cal O}(\alpha)$ corrections of the $M_T$ distribution in the
electron case are shown in Fig.~\ref{fig:four}b. 
To model the detector acceptance, the following $p_T$ and $\eta$ cuts were
imposed in Fig.~\ref{fig:four}:
\begin{equation}
p_T(\ell)>25~{\rm GeV,}\qquad |\eta(\ell)|<1.2, \qquad
\ell=e,\,\mu ,
\label{eq:lepcut}
\end{equation}
\begin{equation}
p\llap/_T>25~{\rm GeV.}
\label{eq:ptmisscut}
\end{equation}
These cuts are similar to the acceptance cuts used by the D\O\
collaboration in their $W$ mass analyses in Run~I. As before, 
uncertainties in the energy and momentum measurements of the charged leptons 
in the detector are simulated in the calculation by Gaussian smearing 
of the particle four-momentum vector. 

The initial state QED-like contribution uniformly increases  the cross 
section by about 1\%. It is largely canceled by the modified weak 
initial state 
contribution. The interference contribution is very small. It decreases
the cross section by about $0.01\%$ for transverse masses below $M_W$, and
by up to $0.5\%$ for $M_T>M_W$. The final state QED-like contribution 
significantly changes 
the shape of the transverse mass distribution and reaches its maximum
effect in the region of the Jacobian peak, $M_T\approx M_W$. As for the 
initial state,
the modified weak final state contribution reduces the cross section by 
about $1\%$, and has no effect on the shape of the transverse 
mass distribution. Since the 
final state QED-like contribution is proportional to $\log(\hat
s/m_\ell^2)$, its size for muons is considerably smaller than for
electrons. The initial state corrections and the interference
contribution are very similar for electron and muon final states. 

\begin{figure}[t]
\centerline{
\epsfysize=2.5in
\epsffile{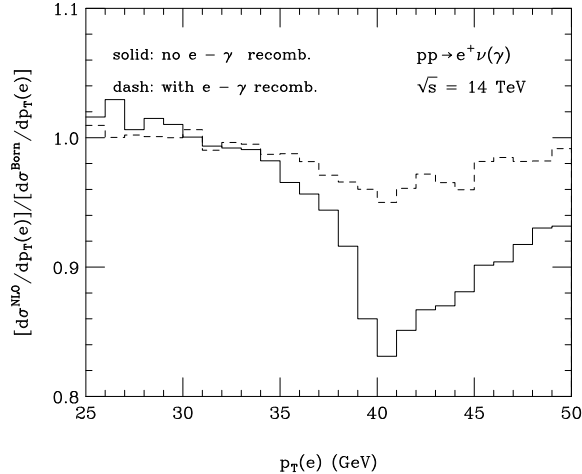}
}
\caption{Ratio of the \protect{${\cal O}(\alpha^3)$} and lowest order
$pp\to e^+\nu(\gamma)$ 
differential cross sections at the LHC as a function of the transverse
momentum of the electron with and without electron -- photon recombination.
The cuts imposed are described in the text.}
\label{fig:five}
\end{figure}

In Fig.~\ref{fig:four}, we have not taken into account the recombination 
of electrons and photons if their opening angle is small. As in $Z$
boson production, when recombination is included, the mass singular 
logarithmic terms are eliminated. This significantly reduces the size of 
the EW corrections. Figure~\ref{fig:five} demonstrates the effect for
the transverse momentum distribution of the electron in 
$pp\to e^+\nu(\gamma)$ at the LHC.
The solid histogram
shows the cross section ratio taking only the transverse momentum and
pseudorapidity cuts of Eqs.~(\ref{eq:lepcut}) and~(\ref{eq:ptmisscut}) 
into account. The dashed histogram displays the result 
obtained when in addition the four momentum vectors are smeared
according to the ATLAS specifications~\cite{ATLTDR}, and electron and
photon momenta are combined if $\Delta
R(e,\gamma)<0.07$~\cite{ATLTDR}. Qualitatively similar results are
obtained for Tevatron energies and CDF and D\O\ lepton identification
criteria. 

As we have seen, final state bremsstrahlung has a non-negligible effect
on the shape of the $M_T$ distribution in the Jacobian peak region. As
in the $Z$ boson case, final state photon radiation shifts the
$W$ boson mass extracted from data to a lower value. In the 
approximate treatment of
the electroweak corrections used so far by the Tevatron experiments,
only final state QED corrections are taken into account; initial state,
interference, and weak correction terms are ignored. Furthermore, the 
effect of the final state soft and virtual photonic
corrections is estimated from the inclusive ${\cal O}(\alpha^2)$ 
$W\to\ell\nu(\gamma)$ width and the hard photon 
bremsstrahlung contribution~\cite{BK}. When detector effects are
included, the approximate calculation leads to a shift of about
$-50$~MeV in the electron case, and approximately $-160$~MeV in the muon
case~\cite{kotwal}. Since only one of the $W$ decay products radiates
photons, the shift in $M_W$ is about a factor~2 smaller than the shift
in $M_Z$ caused by photon radiation. 

\begin{figure}[t]
\centerline{
\epsfysize=3.in \epsffile{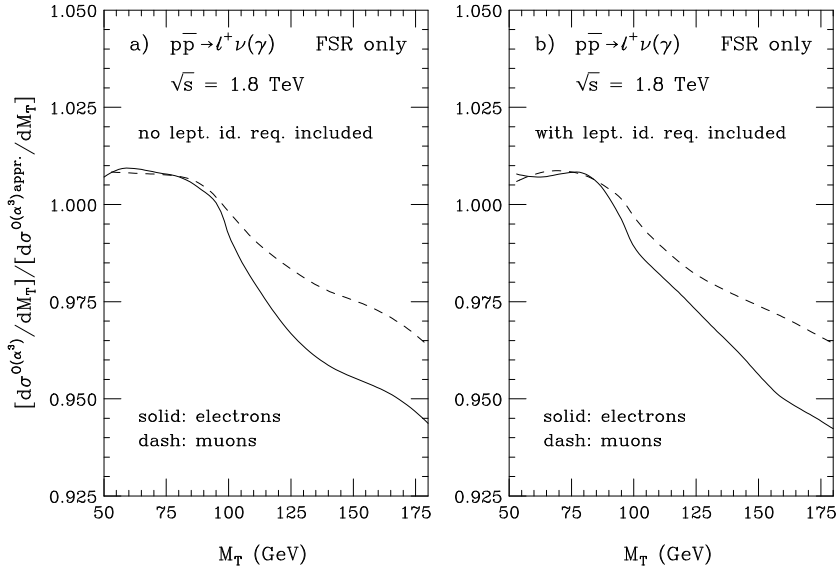}
}
\caption{Ratio of the $M_T$ distributions obtained with the QED-like
final state correction part of Ref.~[\ref{saku}] to the one obtained using
the approximation of Ref.~[\ref{ber}] for
$p\bar p\to\ell^+\nu(\gamma)$ at $\protect{\sqrt{s}=1.8}$~TeV. }
\label{fig:six}
\end{figure}

Initial state and interference contributions do not change the shape of 
the $M_T$ distribution significantly (see Fig.~\ref{fig:four}b) and 
therefore have little effect on the
extracted mass. However, correctly incorporating the final state 
virtual and soft photonic corrections results in a non-negligible 
modification of the shape of the transverse mass distribution for
$M_T>M_W$. For $W$ production at the Tevatron this is demonstrated in
Fig.~\ref{fig:six}, which shows the ratio of the $M_T$ distribution
obtained with the QED-like final state correction part of our
calculation to the one obtained using the approximation of
Ref.~\cite{BK}. 

The difference in the line shape of the $M_T$ distribution between the
${\cal O}(\alpha^3)$ calculation of Ref.~\cite{BKW} and the approximation 
used so far occurs in a region which is important for both the
determination of the $W$ mass, and the direct measurement of the $W$
width. The precision which can be achieved in a measurement of $M_W$
using the transverse mass distribution strongly depends on how steeply
the $M_T$ distribution falls in the region $M_T\approx M_W$. Any change 
in the theoretical prediction of the line shape thus directly influences 
the $W$ mass measurement. From a maximum likelihood analysis the shift
in the measured $W$ mass due to the correct
treatment of the final state virtual and soft photonic corrections is
found to be $\Delta M_W\approx {\cal O}(10~{\rm MeV})$. For the
precision expected in Run~II and for the LHC such a shift cannot be
ignored. 

Two photon radiation has only a modest effect on the shape of the $M_T$
distribution~\cite{BS}. Detailed simulations will be necessary to
determine its effect on the $W$ mass.

The calculation presented in Ref.~\cite{BKW} was carried out in the pole 
approximation, ie. the form factors associated with the modified weak
corrections were evaluated at $\hat s=M_W^2$. This approximation is
valid in the vicinity of the $W$ pole. Away from the resonance region it will 
be important to go beyond the pole approximation. Calculations which
do so are in progress~\cite{BWKD}. Preliminary results~\cite{ewlhc} indicate
that the radiative corrections for $p_T(\ell)>200$~GeV in a full
calculation are up to a factor~2 larger than those calculated in the
pole approximation. This may be important for a measurement of the $W$
width from the high transverse mass tail.

\section{Summary and Outlook}
\label{sec:four}
Accurate theoretical predictions for $W$ and $Z$ boson production are
essential for many important electroweak precision measurements in
future hadron collider experiments, in particular the measurement of the 
$W$ mass and width. In addition, comparison of the $Z$ boson mass and 
width with the values obtained at LEP will help to calibrate
detectors. All these measurements require a detailed understanding of 
the EW radiative corrections. I have described the current status of 
calculations of the ${\cal O}(\alpha)$ EW corrections to $W$ and $Z$
boson production in hadronic collisions. These calculations will be
complete by the time Run~II of the Tevatron is expected to start. Much
more work is required to determine the effect of multiple photon
radiation on the weak boson masses extracted from hadron collider
experiments. 

\section*{Acknowledgements}
This work has been supported by NSF grant PHY-9970703.


\begin{references}
%
\bibitem{lepewk}
D.~Abbaneo {\it et al.} (The LEP Electroweak Working Group),
CERN-EP-2000-016 (January~2000).
%
\bibitem{lephiggs}
P.~Bock {\it et al.} (The LEP Working Group on Higgs Boson searches), 
CERN-EP-2000-055 (April~2000).
%
\bibitem{straes}
A.~Straessner, talk given at the XXXVth Rencontres de Moriond,
``Electroweak Interactions and Unified Theories'', 11 -- 18 March 2000, 
Les Arcs, France. 
%
\bibitem{Tev2000}
H.~Aihara {\it et al.}, in ``Future Electroweak Physics at the Fermilab
Tevatron: Report of the TEV\_2000 Study Group'', edited by D.~Amidei and
R.~Brock, Fermilab-Pub-96/082, 1996, p.~63.
%
\bibitem{LEPWmass}
D.G.~Charlton, hep-ex/9912019, To be published in the proceedings of the
``19th International Symposium on Lepton and Photon Interactions at 
High-Energies'', Stanford, California, 9 -- 14 August 1999. 
%
\bibitem{ewlhc}
S.~Haywood {\it et al.}, hep-ph/0003275, Proceedings of the ``1999 CERN
Workshop on Standard Model Physics (and more) at the LHC'', CERN Yellow
Report CERN-2000-004, eds. G.~Altarelli and M.~Mangano.
%
\bibitem{kotwal}
F.~Abe {\it et al.}  (CDF Collaboration), Phys.\ Rev.\ Lett.\  
{\bf 75}, 11 (1995) and Phys.\ Rev.\  {\bf D52}, 4784 (1995);
T.~Affolder {\it et al.}  (CDF Collaboration), hep-ex/0007044 (July~2000);
S.~Abachi {\it et al.}  (D\O\ Collaboration), Phys.\ Rev.\ Lett.\  
{\bf 77}, 3309 (1996); 
B.~Abbott {\it et al.}  (D\O\ Collaboration), Phys.\ Rev.\  {\bf D58},
012002 (1998), Phys.\ Rev.\  {\bf D58}, 
092003 (1998) and Phys.\ Rev.\ Lett.\  {\bf 80}, 3008 (1998).
%
\bibitem{BK}
\label{ber}
F.A.~Berends and R.~Kleiss, Z.~Phys. {\bf C27}, 365 (1985).
%
\bibitem{BKW}
U.~Baur, S.~Keller and D.~Wackeroth, Phys. Rev. {\bf D59}, 013002 (1999).
%
\bibitem{BKS}
\label{saku}
U.~Baur, S.~Keller and W.K.~Sakumoto, Phys. Rev. {\bf D57}, 199 (1998).
%
\bibitem{BBHSW}
U.~Baur, O.~Brein, W.~Hollik, C.~Schappacher and D.~Wackeroth, in
preparation. 
%
\bibitem{NLOMC}
H.~Baer, J.~Ohnemus, and J.F.~Owens, Phys. Rev. {\bf D40}, 2844 (1989);
W.~Giele and E.W.N.~Glover, Phys. Rev. {\bf D46}, 198 (1992).
%
\bibitem{spies}
J.~Kripfganz and H.~Perlt, Z.~Phys. {\bf C41}, 319 (1988);
H.~Spiesberger, Phys. Rev. {\bf D52}, 4936 (1995); 
A.~de Rujula {\it et al.}, Nucl. Phys. {\bf B154}, 394 (1979). 
%
\bibitem{BS}
U.~Baur and T.~Stelzer, Phys.\ Rev.\  {\bf D61}, 073007 (2000).
%
\bibitem{log}
P.~Ciafaloni and D.~Comelli, Phys.\ Lett.\  {\bf B446}, 278 (1999).
%
\bibitem{KPS}
J.~H.~K\"uhn and A.~A.~Penin, hep-ph/9906545;
J.H.~K\"uhn, A.A.~Penin and V.A.~Smirnov, hep-ph/9912503 and
hep-ph/0005301;
V.~S.~Fadin, L.~N.~Lipatov, A.~D.~Martin and M.~Melles,
Phys.\ Rev.\  {\bf D61}, 094002 (2000);
M.~Melles, hep-ph/0004056 and hep-ph/0006077;
W.~Beenakker and A.~Werthenbach, hep-ph/0005316.
%
\bibitem{dw}
W.~Hollik and D.~Wackeroth, Phys. Rev. {\bf D55}, 6788 (1997).
%
\bibitem{ATLTDR}
ATLAS Collaboration, ``ATLAS Detector and Physics Performance'',
CERN/LHCC/99-15 (1999).
%
\bibitem{BWKD}
U.~Baur and D.~Wackeroth, in preparation; S.~Dittmaier and M.~Kr\"amer,
in preparation.
%
\end{references}
\end{document}